\documentclass[final]{cvpr}

\usepackage{times}
\usepackage{epsfig}
\usepackage{graphicx}
\usepackage{amsmath}
\usepackage{amssymb}
\usepackage{subfigure}
\usepackage{overpic}

\usepackage{bm}
\usepackage{algorithm}

\usepackage{algorithmicx}
\usepackage{algpseudocode}

\usepackage{enumitem}
\setenumerate[1]{itemsep=0pt,partopsep=0pt,parsep=\parskip,topsep=5pt}
\setitemize[1]{itemsep=0pt,partopsep=0pt,parsep=\parskip,topsep=5pt}
\setdescription{itemsep=0pt,partopsep=0pt,parsep=\parskip,topsep=5pt}

\usepackage[pagebackref=true,breaklinks=true,colorlinks,bookmarks=false]{hyperref}

\def\cvprPaperID{159} 
\def\confYear{CVPR 2020}

\graphicspath{{figures/}}

\newcommand{\LTM}{ A}
\newcommand{\gcut}{\mathfrak{g}}
\newcommand{\slice}{\mathbf s}
\newcommand{\matX}{ X}
\newcommand{\matY}{ Y}
\newcommand{\matM}{ M}
\newcommand{\vOne}{\mathbf e}

\newcommand{\refSec}[1]{Section \ref{#1}}
\newcommand{\refTab}[1]{Table \ref{#1}}
\newcommand{\refFig}[1]{Figure \ref{#1}}

\newcommand{\Ignore}[1]{}

\newcommand{\NewFigure}[3]{
  \begin{figure}[thb]
  \centering
  {#2} 
  \caption{#3}
  \label{#1}
  \end{figure}
}

\newcommand{\NewWideFigure}[3]{
  \begin{figure*}[thb]
  \centering
  {#2} 
  \caption{#3}
  \label{#1}
  \end{figure*}
}

\newcommand{\NullFigure}[3]{}

\newcommand{\NewTable}[3]{
  \begin{table}[thb]
  \centering
  {#2}
  \caption{#3}
  \label{#1}
  \end{table}
}

\title{Sparse Sampling and Completion for Light Transport in VPL-based Rendering}

\author{papers 0578}
\author{Yuchi Huo ~~~~~ Rui Wang ~~~~~ Xinguo Liu ~~~~~ Hujun Bao 
}

\begin{document}

\maketitle

\begin{abstract}
The many-light formulation provides a general framework for rendering various illumination effects using hundreds of thousands of virtual point lights (VPLs). To efficiently gather the contributions of the VPLs, lightcuts and its extensions cluster the VPLs, which implicitly approximates the lighting matrix with some representative blocks similar to vector quantization. In this paper, we propose a new approximation method based on the previous lightcut method and a low-rank matrix factorization model. As many researchers pointed out, the lighting matrix is low rank, which implies that it can be completed from a small set of known entries.

We first generate a conservative global light cut with bounded error and partition the lighting matrix into slices by the coordinate and normal of the surface points using the method of lightslice. Then we perform two passes of randomly sampling on each matrix slice. In the first pass, uniformly distributed random entries are sampled to coarsen the global light cut, further clustering the similar light for the spatially localized surface points of the slices. In the second pass, more entries are sampled according to the possibility distribution function estimated from the first sampling result. Then each matrix slice is factorized into a product of two smaller low-rank matrices constrained by the sampled entries, which delivers a completion of the lighting matrix. The factorized form provides an additional speedup for adding up the matrix columns which is more GPU friendly. Compared with the previous lightcut based methods, we approximate the lighting matrix with some signal specialized bases via factorization. The experimental results shows that we can achieve significant acceleration than the state of the art many-light methods.
\end{abstract}

\begin{figure*}
    \centering
\begin{tabular} {ccc}
     \includegraphics[width=2.2in]{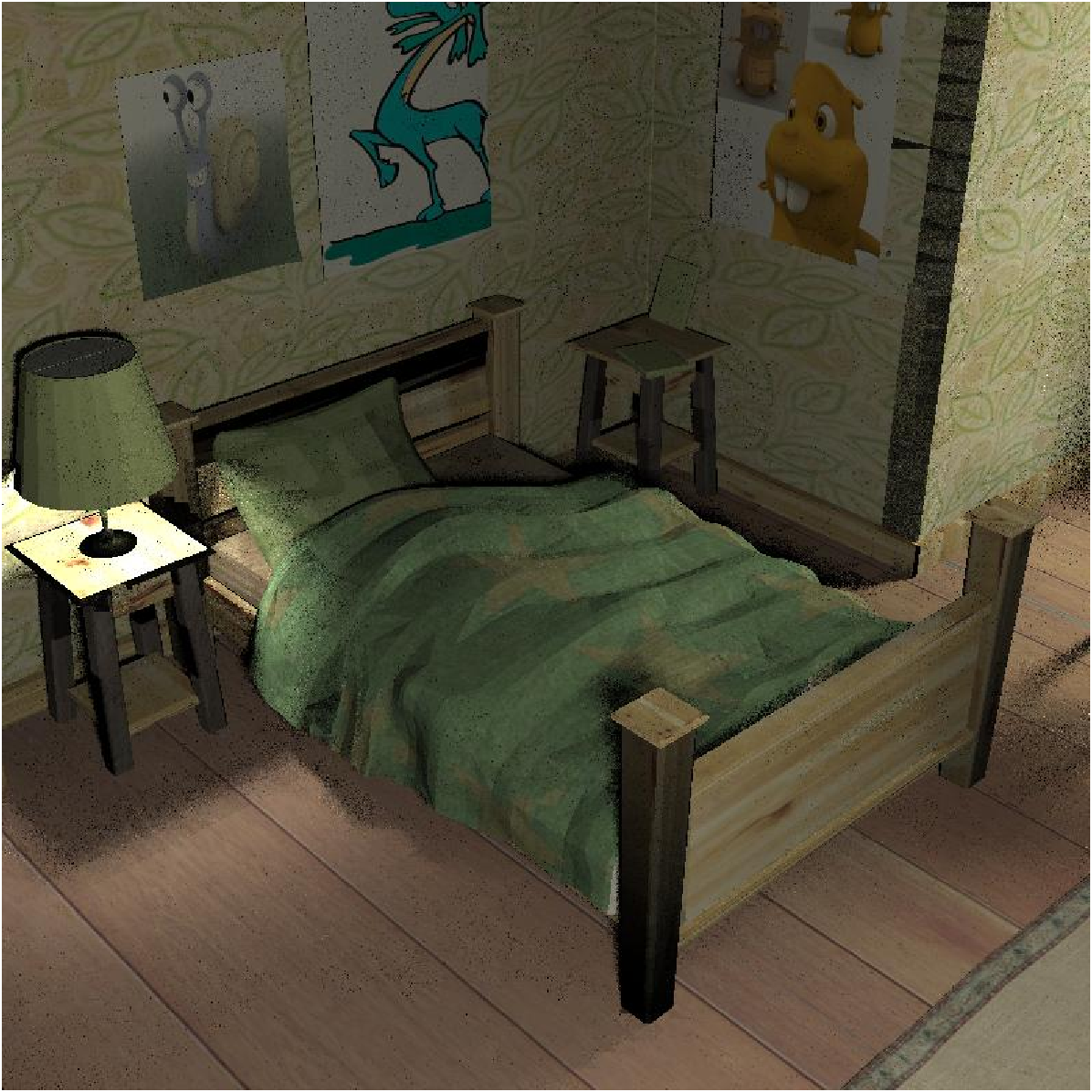}&
     \includegraphics[width=2.2in]{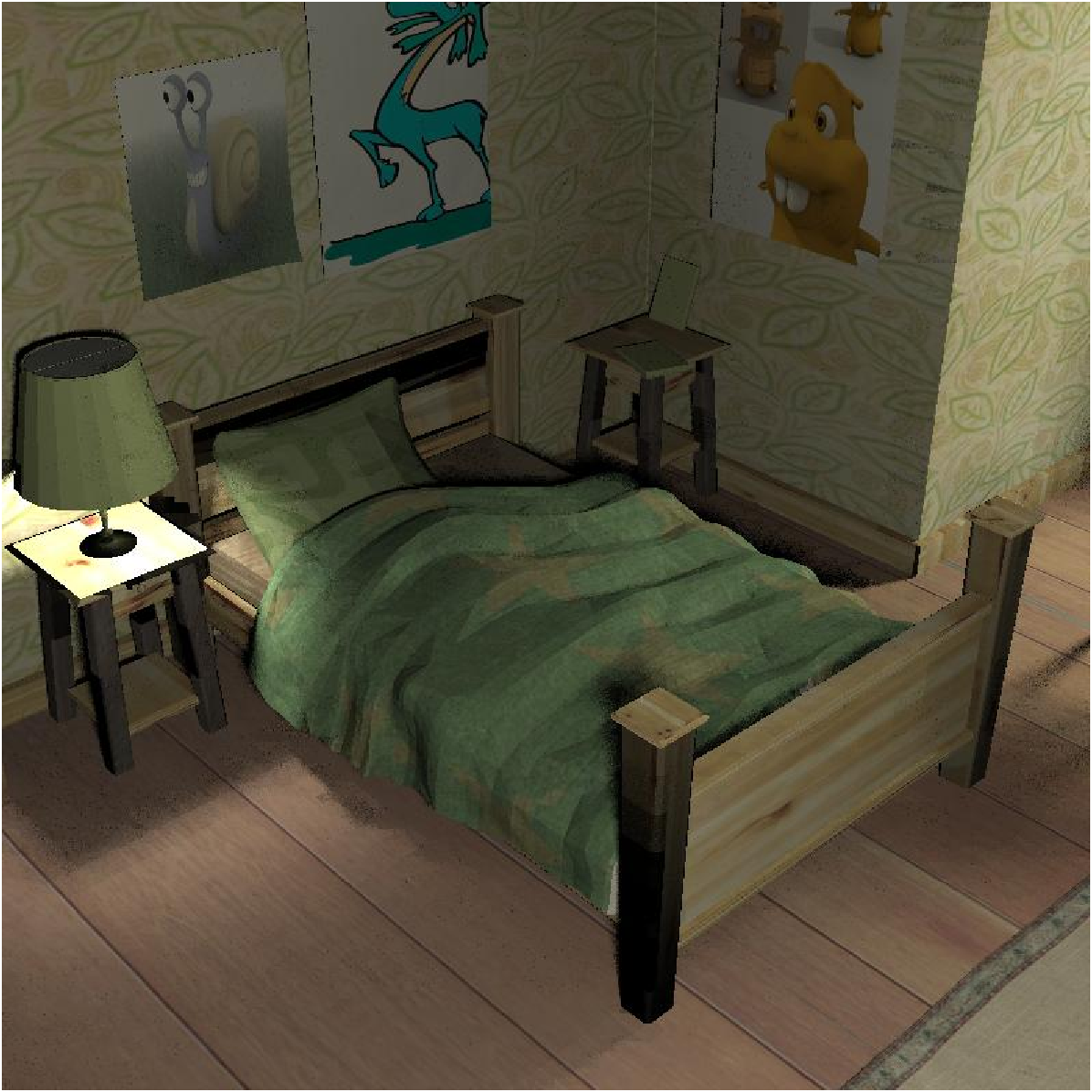}&
     \includegraphics[width=2.2in]{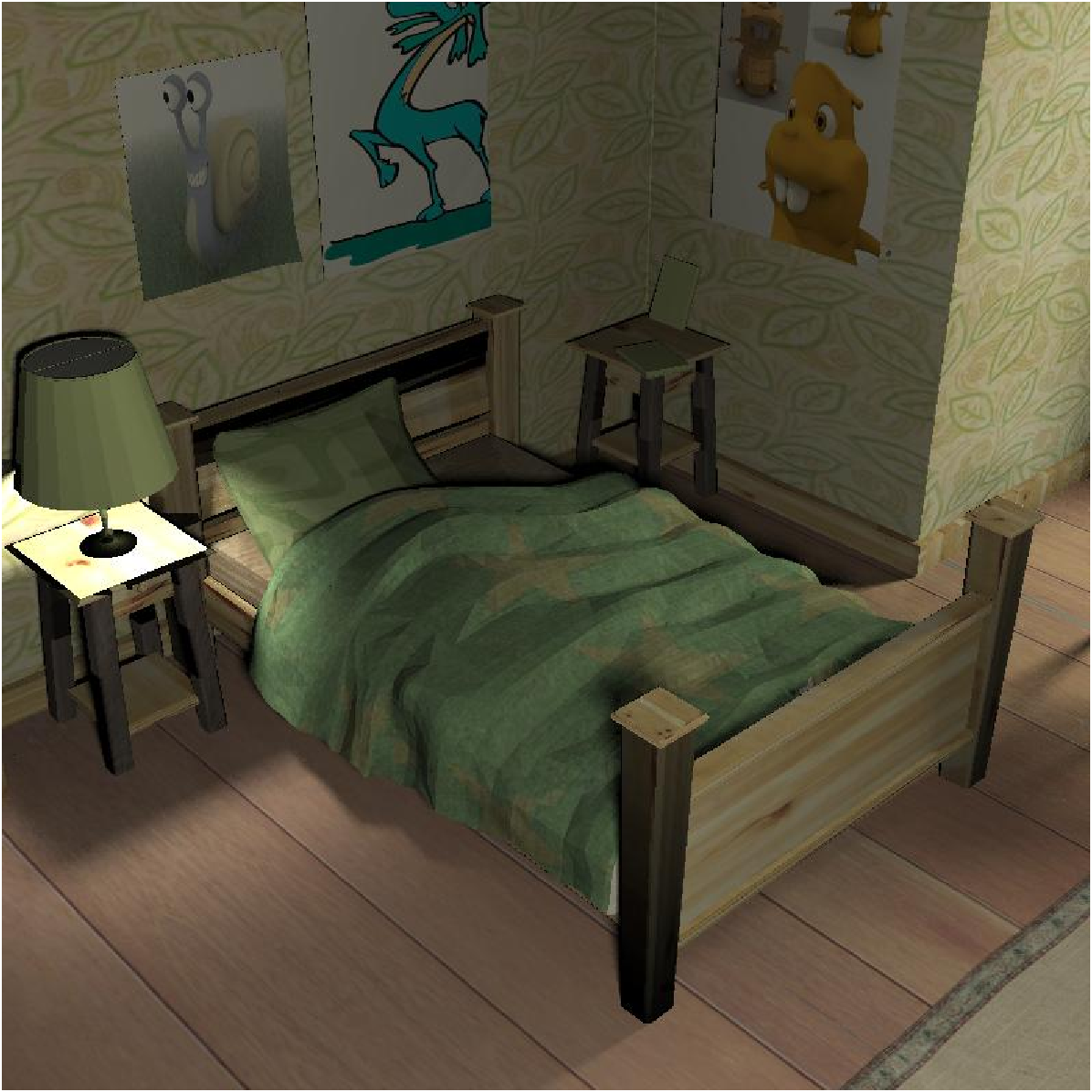}\\
     1\% sampling &
     5\% sampling &
     10\% sampling \\
    \end{tabular}
    \caption{The above images are rendered by the proposed methods with various sampling rate.}
    \label{fig:teaser}
\end{figure*}

\section{Introduction}


Rendering realistic images with global illumination evolves integrating the rendering equation, which is a recursively defined expression because the incoming lighting is to be determined by the same equation. Various efforts are attributed to high-quality rendering \cite{wang2013gpu,wangimplementation,huo2015matrix,huo2016adaptive,huo2020spherical,cho2021weakly,fan2021real,huo2021survey,huo2020adaptive,huo2022extension}, or to achieve a balance between quality, performance, and flexibility \cite{kim2020single,li2021multi,an2021hypergraph,park2021meshchain,zhang2021powernet,li2020automatic}. Keller~\cite{Keller:InstantRadiosity} introduced an intermediate representation, called virtual point light (VPL), to break the recursive rendering equation, yielding a fast realistic rendering algorithm and inspiring a number of followups called VPL-based rendering. VPL-based rendering first generates a set of VPLs by tracing rays from the real light sources, and then renders the scene by accumulating the direct illumination results of the VPLs. The rendering results can be also expressed as summing the columns of a matrix, called the lighting matrix, where each row corresponds to a pixel and each column corresponds to a light. However, the cost of computing the lighting matrix will become prohibitively expensive when there are hundreds of thousands of VPLs required for faithful approximation of the lighting effects, which is unfortunately very common in high quality rendering applications.

Recently, many extensions and accelerations of the VPL-based rendering have been developed, among which a class of work, called the lightcut based method, focused on efficiently solutions based on cutting/clustering ~\cite{Walter:lightcuts05,Walter:lightcuts06,Hasan:MatrixRowColumn,Ou:2011:LightSlice}. The lightcut method~\cite{Walter:lightcuts05} showed that, while the number of lights is huge, they can be aggressively clustered as some representatives without degrading the perceptual quality of the result image. The followup work, row-column sampling method \cite{Hasan:MatrixRowColumn}, finds the same set of representative lights for the whole image for further speedup, and the light slice method \cite{Ou:2011:LightSlice} refine the clusters for preserving the details by slicing the image into smaller blocks.

From the view of the lighting matrix, the lightcut based methods approximate the lighting matrix with some vectorized blocks. In this paper, we propose to approximate the light transport matrix by a product of two low-rank factor matrices. As the lightslice method did, we first partition the matrix into slices and generate a lightcut for each slice. Then, instead of computing all entries, we sparsely sample the corresponding sub matrix for each slice and complete it using a low-rank factorization method designed for sparse matrix\cite{xu:11:alternating}. In order to capture the matrix structure with as few as possible samples, we design an importance function the lights and a possibility distribution function for sampling the lighting matrix. As shown in \refFig{fig:teaser}, the lighting matrix can be faithfully recovered using $10\%$ sampled entries for rendering the final images. As an byproduct benefit, the low-rank production form allows for faster integration to get the final image, since we can sum the columns of lighting matrix without explicitly recovering the matrix.

The contributions of this work are:
\begin{itemize}
  \item Introduce a sparse factorization method for the lighting matrix in many-light problem.
  \item Propose a high quality lightcut generation algorithm using sparsely sampled lighting and visibility information.
  \item Propose an efficient, GPU friendly sampling and completion algorithm for the lighting matrix.
\end{itemize}

\section{Related Work}

We first review some work on virtual point light, lightcut and some realistic rendering methods related to them, then review some work on matrix factorization.

\textbf{Many-light methods}~~Keller \cite{Keller:InstantRadiosity} introduced Instant Radiosity (IR) method, which is the first realistic rendering algorithm using virtual point lights. It traces a number of virtual point lights (VPLs) from light sources, and then computes the indirect global illumination in the scene as the direct illumination of the virtual point lights. This idea is very attractive, and has inspired many followup works, which manipulate the rendering equation in a many-light framework based on VPL. A key issue in many-light framework is the expensive cost dealing with a large number of lights, which is an order of millions and even more in high quality realistic rendering. Walter et al.~\cite{Walter:lightcuts05} introduced a lightcut method to reduce the pixel-light computation cost from linear to sub linear by clustering the lights into a few hundreds of representatives. The multidimensional lightcuts method~\cite{Walter:lightcuts06} extended lightcut to handle high dimensional integrations for rendering volume scattering, depth of field or motion blur, etc. Hasan et al.~\cite{Hasan:MatrixRowColumn} proposed a sampling method to generate a global light cut optimized for the whole image, which greatly reduce the cost for cutting the light tree for every pixels. To further exploit the spatial coherence and adapt the image details, Ou et al.~\cite{Ou:2011:LightSlice} partition the image into slices and refine the global lightcut for each slice. Note that visibility test is the most expensive step in computing the light transport entries. The work of Ritschel et al.~\cite{Ritschel:ImperfectShadowmap} shows that approximate visibility test with imperfect shadow map can effectively reduce the cost for indirect illumination.

An limitation of VPL based rendering lies in difficulties for handling glossy materials. In order to handle glossy materials and subsurface scattering, Walter et al.~\cite{Walter:2012:BL} introduced a new intermediate representation, called virtual point sensors, to compensate the bias due to VPL clamping. Hasan et al.~\cite{Hasan:2009:VSL} extended VPL to virtual spherical light (VSL) to avoids the clamping bias. However it may bring blurry artifacts on sharp features. Davidovic et al.~\cite{Davidovic:2010:CGL} introduces local VPLs to compensate the loss of clamping and offers better approximation for sharp glossy reflections.

\textbf{Path tracing methods}~~Since the pioneer work of Whitted~\cite{Whitted1980}, path tracing has been standard framework for simulating global illumination. Bidirectional ray tracing combines the light path tracing and eye path tracing to inherit the specific advantages of path tracing as well as light tracing~\cite{Veach:MonteCarlo}. Monte Carlo path tracing is unbiased, but suffers from low convergence speed and perceptible noise. Photon mapping pioneered by Jensen~\cite{Jensen:PhotonMapping} traces photons in the scene and then caches and reuses them for final gathering of paths from eye. However, photon mapping usually requires a large number of photons to reduce noise and reproduce sharp details, especially for glossy materials and caustics. The memory requirement is proportional to the number of photons, which is thus very high. Progressive photon mapping~\cite{Hachisuka:2008:ASIA} alleviates this problem by multi-pass processing.

\textbf{Cache-based methods}~~Irradiance caching~\cite{IrradianceCaching} progressively caches diffuse irradiance samples as an octree, and reuses them along the computation. Radiance caching \cite{RadianceCaching} extends it by recording directional radiance using spherical harmonics. Bala et al. \cite{Bala:RadianceInterpolant} presented a general approach to exploiting both spatial and temporal coherence of rays for radiance interpolation. Okan Arikan et al.~\cite{Arikan:FastGI} introduced a fast approximation to global illumination by decomposing radiance fields into far- and near-field components, which are computed separately to improve efficiency. For glossy interreflections, Krivanek et al.~\cite{Krivanek:SpatialCache} proposed a spatial directional cache method for glossy to glossy reflections.  The GPU accelerated irradiance caching by Wang et al.~\cite{Wang:2009:EGA} handles only diffuse interreflections.

\textbf{Matrix factorization}~~ Matrix recovery has a wide range of applications in image processing, graphics, vision and machine learning. A recovery of a partial matrix is a specific choice of values for the unspecified entries. Given a few assumptions on the nature of the matrix, e.g. the matrix is low rank and the observed entries are randomly sampled, the missing entries can be recovered by various algorithms \cite{Candes:2012:matrix}. In this paper, we chose an nonnegative factorization method \cite{xu:11:alternating}to recover the sparsely sampled lighting matrix. There have been a few applications of matrix factorization in rendering, e.g. separable BRDF\cite{kautz:99}, environment mapping\cite{McCool:01:homomorphic}, precomputed radiance transfer~\cite{sloan:cpca}. However, none of them touched sparsely sampled matrix with a lot of missing entries.

\section{Overview}
This section gives an overview of the presented rendering algorithm. We adopt the matrix formulation \cite{Hasan:MatrixRowColumn,Ou:2011:LightSlice} for the many-light problem to describe the sampling and rendering algorithms.

Suppose that there are $m$ surface points each of which corresponds to a pixel in the rendered image, and $n$ virtual point lights generated for rendering the scene. Let $\LTM$ be the lighting matrix, where each entry $\LTM(i,j)$ represents the illumination contribution from light $j$ to surface point $i$. Then the rendered image can be expressed as $\Sigma_\LTM$, the sum of the columns of matrix $\LTM$.

Since the number of the virtual point lights is very high, naively computing all of the entries in matrix $\LTM$ is impractical. The lightcut based methods cluster the virtual point lights into some representative lights, which merges the columns corresponding to the lights in the same cluster, yielding a reduced lighting matrix. In this paper, we present new techniques to improve the clustering procedure, and further save the computation by sparsely sampling the reduced lighting matrix.

Given the VPLs in the scene and the surface points to be shaded, our rendering algorithm proceeds in 5 main steps as follows:
\begin{enumerate}
  \item \emph{Global Lightcut}: build a light tree for the VPLs, and generate a global cut $\gcut$ of the light tree.

      We adopt the lightcut method in \cite{Walter:lightcuts06} to generate the global lightcut. The global lightcut is error-bounded for the whole image. It is also conservative, because the visibility function is overestimated. In our experiments, the global lightcut contains about $800\sim900$ nodes of lights.

  \item \emph{Matrix Slicing}: partition the matrix into slices by clustering the surface points according to their coordinates and normal.

      Considering both efficiency and stability, smaller sized and lower ranked matrices are preferred for low-matrix factorization methods. The original lighting matrix $\LTM$ usually have several hundreds of thousands of rows, and several hundreds of columns corresponding to the node number in the lightcuts, which makes matrix $\LTM$  difficult to factorize. Therefore, we need to partition the matrix into small ones. We adopt the lightslice method \cite{Ou:2011:LightSlice} to partition the matrix, and set the slice size to be about 800 pixels. As many research pointed out, the lighting matrix of nearby pixels has much lower rank \cite{Mahajan:2007,Ou:2011:LightSlice}. Therefore, it is beneficial to perform low-rank factorization on smaller slices instead of on the whole image.

  \item \emph{Light Coarsening}: for each slice, coarsen the global lightcut $\gcut$ to merge more light node.

      The surface points in each slice are more closely located, then more lights are likely to be distant and can be merged safely without introducing perceptible artifacts. Therefore, we can cut more nodes in the light tree for each slice, so as to further save computational cost. We design a random sampling method and a cost function for coarsening the global lightcut. The details about light coarsening is presented in \refSec{sec:coarsen}.

  \item \emph{Matrix Sampling and Completion}: for each matrix slice, sample a sparse set of entries and perform a low-rank factorization.

      Performing low-rank analysis via SVD on the lighting matrix is mentioned by Ha\v{s}an et al.\cite{Hasan:MatrixRowColumn}. However, naive SVD is impractical for efficient rendering, since the lighting matrix takes long time to compute. Instead, we sample a small percentage of entries in the lighting matrix, and then recovery it by performing low-rank factorization with the known entries as constraints.
      The detail of the sampling and factorization procedure is presented in \refSec{sec:completion}.

  \item \emph{Image Rendering}: At last, the final image is rendered slice by slice.

      Consider a matrix slice $\slice$, and let $\matM$ be the lighting matrix of slice $S$, and its factorization result be $\matM = \matX\cdot\matY$. Then the rendered image restricted on slice $\slice$ can be rewritten as follows:
      $$
      I(\slice) = \matM\cdot\vOne=\left(\matX\cdot\matY\right)\cdot\vOne =\matX\cdot\left(\matY\cdot\vOne\right),
      $$
      where $\vOne$ is a 1D column vector whose elements are all $1$s.
      Therefore, the recovered matrix slices are not necessarily stored for producing the final image, as long as the factor matrices $\matX$ and $\matY$ are kept.
\end{enumerate}

The methods in step 1 and 2 are adopted from the matrix row-column sampling paper \cite{Walter:lightcuts06} and the light slice paper \cite{Ou:2011:LightSlice}, where the readers can find the details. Since step 5 is straight forward, we will present the methods for step 3 and 4 in the next two sections.

\section{Light Coarsening} \label{sec:coarsen}

Consider two sibling leaf nodes, $L_a$ and $L_b$, in the conservative global lightcut $\gcut$, and let $L_f$ be the parent node of $L_a$ and $L_b$. By the definition of lightcut, we have:
$$
I_f = I_a + I_b,
$$
where $I_{a/b/f}$ be the intensity of light node $a/b/f$. Without loss of generality, suppose that $L_f$ is the same as $L_a$ except their intensity values.

To determine if light $L_a$ and $L_b$ can be merged for rendering slice $\slice$, we randomly select some pixels (rows) inside slice $\slice$, and compute the lighting results by light $L_a$ and $L_b$. Since brighter light usually has more impact on the final image, we set the number of the random pixels to be linearly proportional to the total light intensity, i.e. $I_f$, to take more care on significant lights. Once the sampling number is determined, we choose the pixels uniformly distributed inside the image block corresponding to the current slice.

Let $\zeta_f$ be the set of selected pixels, $V_a$ and $V_b$ be vectors of the lighting results by lights $L_a$ and $L_b$ on pixels $\zeta_f$. Then, the vector of the lighting results by light $L_f$ is $V_f = V_a + V_b$. When we merge lights $L_a$ and $L_b$ to coarsen the light tree, the vector of the lighting result is computed as: $V_a\frac{I_f}{I_a}$ (recall that $L_f$ is the same as $L_a$ except intensity value). Therefore, for pixels in $\zeta_f$, the error of cutting light node $L_a$ and $L_b$ is:
$$
\varepsilon(L_f) = \left\| (V_a+V_b)-V_a\frac{I_f}{I_a} \right\|_{\infty} = \left\| V_b-V_a\frac{I_b}{I_a} \right\|_{\infty}.
$$
To take into account the accumulation error, we define the following cost function for cut the light tree at node $L_f$:
\begin{equation} \label{eq:coarsen:error}
cost(L_f) = \varepsilon(L_f) + cost(L_b),
\end{equation}
and we define the cost function as $0$ for the leaf nodes in the global light tree $\gcut$.

Now, we can repeatedly cut two sibling leaf nodes as long as the cost function is less than a prespecified error bound. When proceed with this coarsening procedure, the following case must happen: the two sibling leaf nodes in testing are the parent nodes of two previous coarsening operations. In this case, instead of set $\zeta_f$ as randomly selected pixels, we set $\zeta_f$ to be the following union:
$$
\zeta_f = \zeta_a \bigcup \zeta_b.
$$
The union method in above has two benefits. First, it allows for reusing the lighting results computed in the previous coarsening operations. Second, it gives the light at upper level more sampling, which is naturally consistent with our idea to set the number of sampled pixels to be proportional to the intensity of the merged light. Though the sampling number is doubled when the coarsening operation proceeds to the upper level, the total number is bounded, since the upper level has only about half number of nodes than the next level.

We can also perform the coarsening procedure until a specified number lights is reached. For this purpose, we order the pairs of sibling nodes by their cost functions, then proceed the coarsening from the pair with the least cost.

\textbf{Coarsening v.s. Refinement} In contrast, the lightslice method\cite{Ou:2011:LightSlice} generate the per slice lightcut by refining a coarse global lightcut. It is not easy to tell which one is more efficient than the other one, because they have different sampling strategy for evaluating the cost.

Since our cost function is defined recursively for taking into account the accumulation error, it is not suitable for refinement. But there is no obviously benefit of refinement v.s. coarsening, even if we can drop the accumulation error term. Therefore, we preferred coarsening rather than refinement in this work. However, it is worth to point out, when the initial lightcut to start with is sufficiently close to the optimal lightcut, it would be better to support lightcut optimization in both directions by refinement and coarsening, such that we can take advantage of the coherence among the nearby slices to save computation. We leave this point as a future work.

\section{Matrix Sampling and Completion} \label{sec:completion}
After generating the lightcuts, we have actually sparsely sampled the lighting matrices for all slices in the meanwhile. There are two choice to complete the light matrices. One choice is to directly compute them, as the lightslice method did \cite{Ou:2011:LightSlice}. The other one is to complete them from the sparely sampled values, since they have been proven to be low rank matrices\cite{Mahajan:2007}. We take the second choice.

However, our experiments showed that the random samples computed in the light coarsening step is too sparse to faithfully recovery the lighting matrix. Therefore, we need sample more entries for completion.

Again, consider a slice $\slice$ with $m'$ pixels and $n'$ representative lights. Let $\matM\in R^{m'\times n'}$ be the corresponding lighting matrix. Before sampling the lighting matrix $\matM$, we need a \emph{possibility distribution function} (\emph{pdf}) for determine the indices of the sampling entries. A naive one is uniform distribution. But we would like to design better pdf for sampling, since we have already obtained a very sparse observation of the lighting matrix in the previous step for light coarsening. To capture the lighting matrix as much as possible with prescribed number of samplings, we prefer to sample important entries. Though it is not clear what kinds of entries are important, the entries corresponding to important light and important pixel are likely to be important by intuition.

Let $f(i)$ be the importance function of pixel $i$ and $g(j)$ be the importance function of light $j$, then we define the \emph{pdf} on the lighting matrix as follows:
\begin{equation}
pdf(i,j)\propto f(i)g(j).
\end{equation}

A light can be regarded as important if it produces various lighting results over all pixels, neither consistent dark nor consistent bright. Therefore, we can measure its importance by the variance of its lighting results. Let $C_j$ be the observed column vector corresponding to light $j$ in the lighting matrix. Then, we define the following importance function for light $j$:
 $$
 g(j) = \max(C_j)-\min(C_j).
 $$
For pixel importance function, we simply treat all pixel equally, i.e. define $f(i)=1$ for all pixels, because no useful information is found from the observed lighting matrix for now.

Therefore, to sample a new matrix element, we first select the index for the light (or column) by the light importance function, then select the index for the pixel (or row). If the element of the sampled indices is empty, then compute the lighting results of the corresponding light-pixel pair. The sampling procedure is repeated until a prescribed number of elements is reached.

To complete the lighting matrix with the sparsely sampled entries, we choose a nonnegative matrix factorization method \cite{xu:11:alternating} based on the algorithm of alternating direction method of multipliers. A brief introduction of the iteration procedure can be found in the appendix. This factorization method gives high-quality solutions within a few iterations. In experiments, we stop the iteration when the error is below a specified tolerance, or the iteration exceeds a specified number. We set the maximum iteration as 100.

After factorization, we have a completed lighting matrix for slice $\slice$:
$$
\matM = \matX\cdot \matY,
$$
where $\matX\in R^{m'\times q}$ and $\matY\in R^{q\times n'}$ are both low rank ($\leq q$) factor matrices. $q$ is user specified parameter as a rank estimation for the input matrix. Smaller $q$ is faster for iteration, but may lead to large factorization error, and vice versa. We experimented with several values $q=8, 16$ and $32$, and found that $q=16$ delivers satisfactory results regarding both efficiency and accuracy. We implemented the factorization method in graphics hardware using Nivdia's Cuda BLAS to accelerate the matrix computation of the iterations.

\NewWideFigure{fig:ref} {
  \begin{tabular} {cc}
     \includegraphics[width=2.2in]{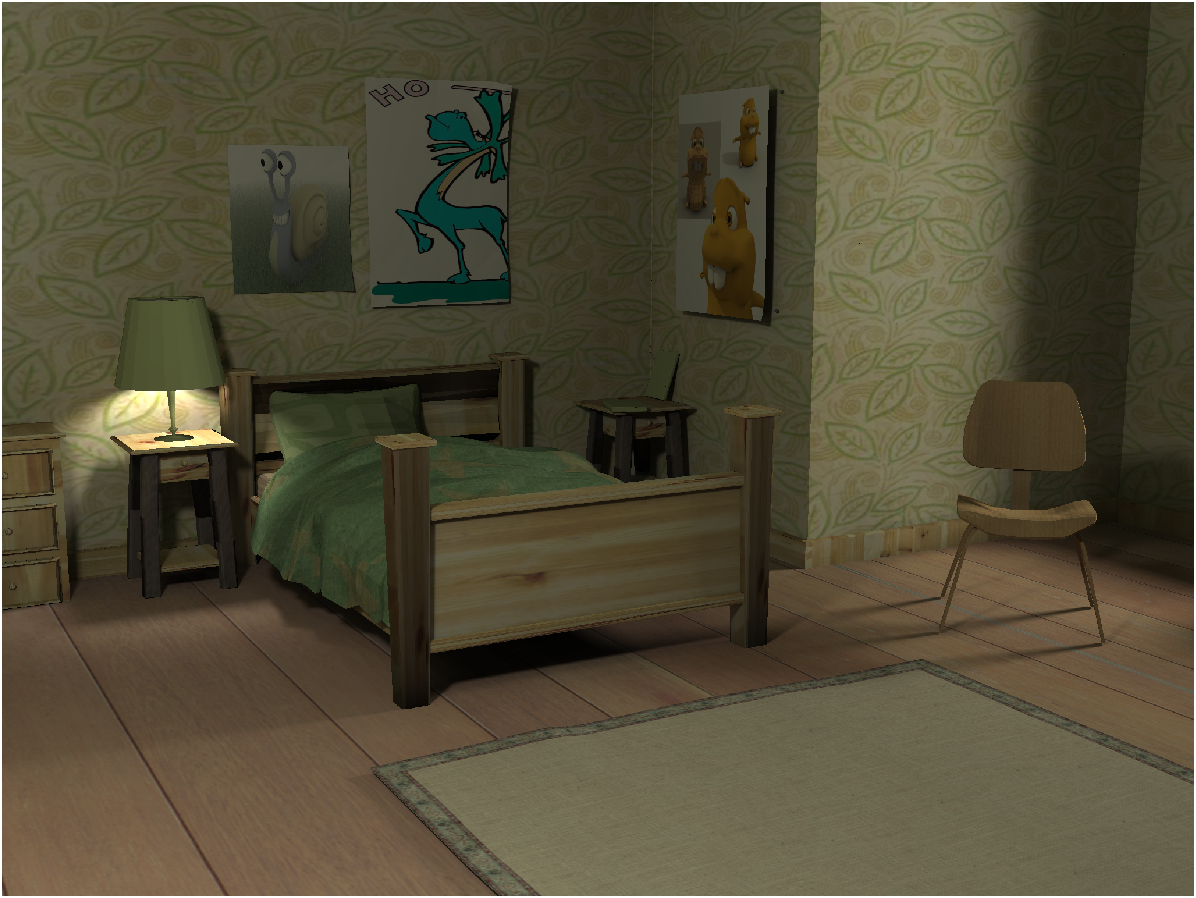} &
     \includegraphics[width=2.2in]{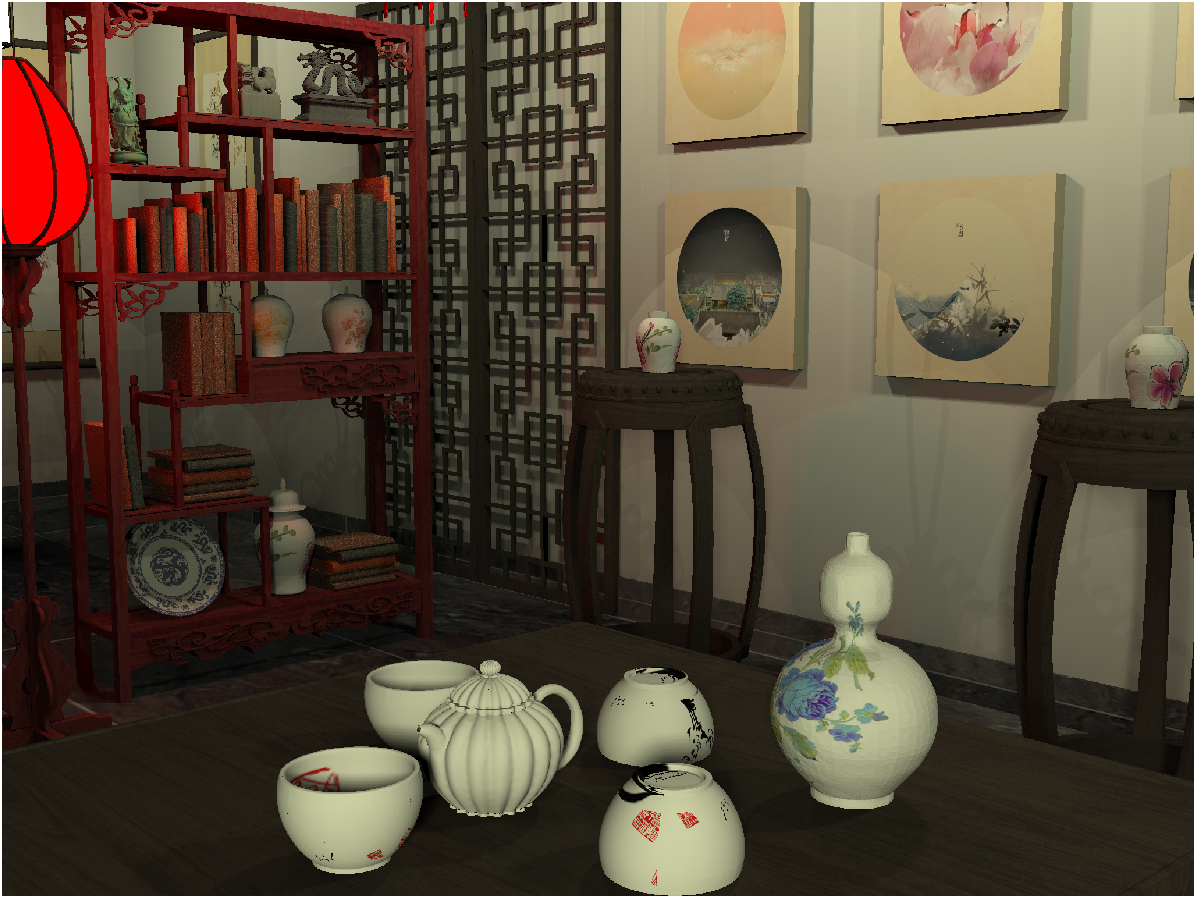} 
     \\
     (a) Bedroom & (b) Tearoom 
  \end{tabular}
} {Some testing scenes. The images are rendered by brute-force rendering using all of the VPLs, and will be used as reference for later evaluation. }

\section{Implementation and Experimental Results}
We have implemented the proposed algorithms, and accelerated them with graphics hardware using Nvidia's CUDA sdk.
In step 1, the global lightcut is generated by splitting the light tree, and the splitting test is parallelized for all nodes level by level in GPU.%
In step 2, the surface points are represented by 6D vectors (coordinates and normal), and are recursively split into a binary tree. We parallel the computation on the tree nodes in GPU. %
In step 3, the coarsening operation is implemented in GPU by parallelizing the computation for all pairs of leaf nodes.%
In step 4, the matrix factorized is implemented with Cuda BLAS.%
The final rendering in step 5 is done GPU too. The presented algorithms frequently require shooting rays to compute the pixel-light visibility. We implement a ray-casting method in GPU accelerated by SBVH\cite{sbvh}. To compare with the previous lightcut based methods, we integrate the CPU implementation code by the authors of the lightslice work\cite{Ou:2011:LightSlice} (which is public available from http://www.cs.dartmouth.edu/~fabio/publication.php?id=lightslice11) into our rendering system, and accelerate the visibility computation procedure using our SBVH implementation in GPU.

In the following of this section, we will present some rendering results on 3 testing scenes, and compare our algorithms with the state of the art lightcut based methods. The reference images of the testing scenes are shown in \refFig{fig:ref}. The statistics and timings of our experiments are summarized in \refTab{tab:scenes}.

\newcommand{\LImg}{1024\times 768}
\newcommand{\MImg}{800\times 600}
\newcommand{\SImg}{512\times 512}

\NewTable{tab:scenes}{
\begin{tabular}{|lccc|}
\hline
Scene     & Bedroom   & Tearoom   & Museum  \\
\hline
\#Tris     & 88K       & 3115K     & 1604K   \\
\#VPLs    & 50K       & 231K      & 165K    \\
$m$       & $\LImg$   & $\LImg$   & $\MImg$ \\
\#Nodes   & 569        & 345        & 609      \\
Time(s)   & 8        & 9        & 18      \\
\hline
\end{tabular}
}{Summary of the scene size and the rendering performance. \#Tris is the number of triangles in the scenes, \#Nodes is the averaged number of the leaf nodes in the coarsened light tree. Time is measured in seconds. }

\NewFigure{fig:museum} {
  \begin{tabular} {c}
     \includegraphics[width=3in]{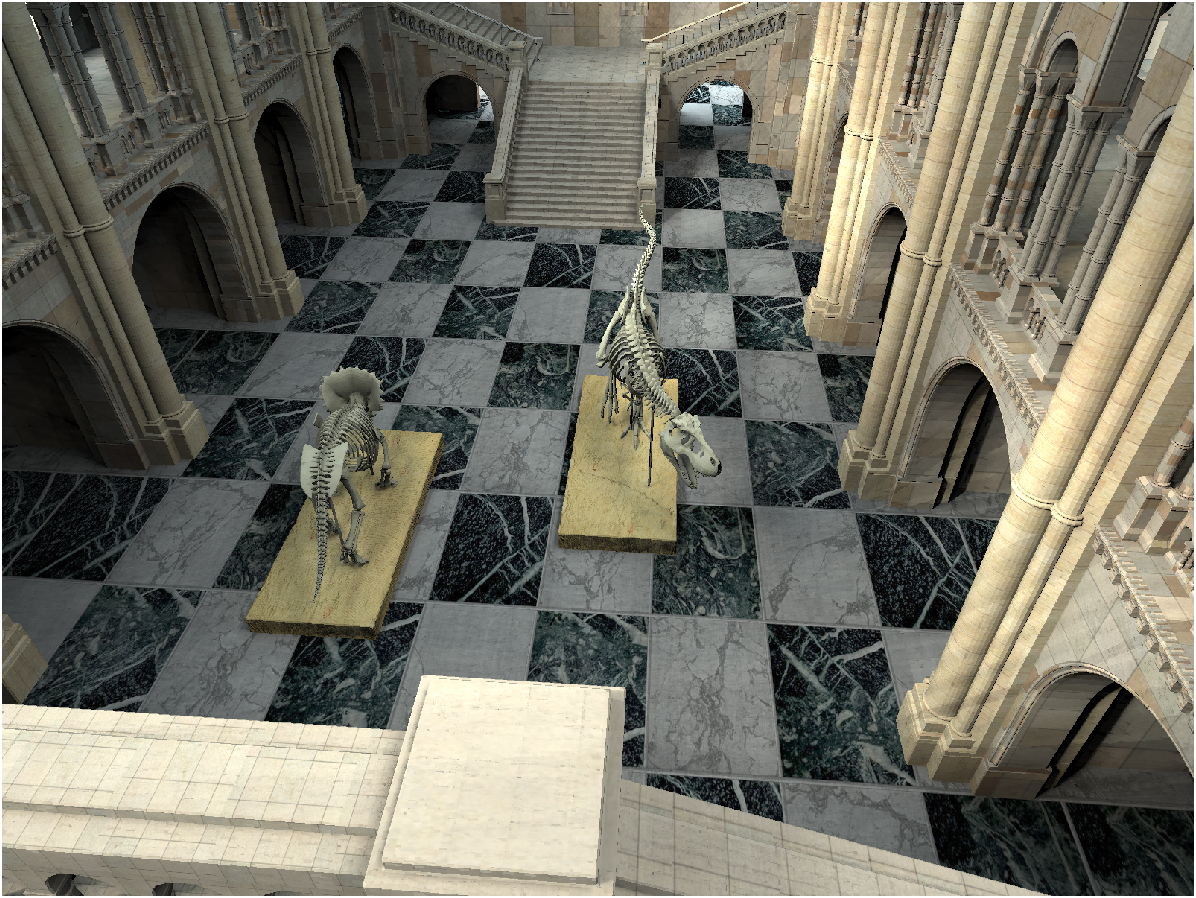}\\
  \end{tabular}
} {The Museum scene rendered by our method. }

\NewWideFigure{fig:f1} {
  \begin{tabular} {ccc}
     \includegraphics[width=2.2in]{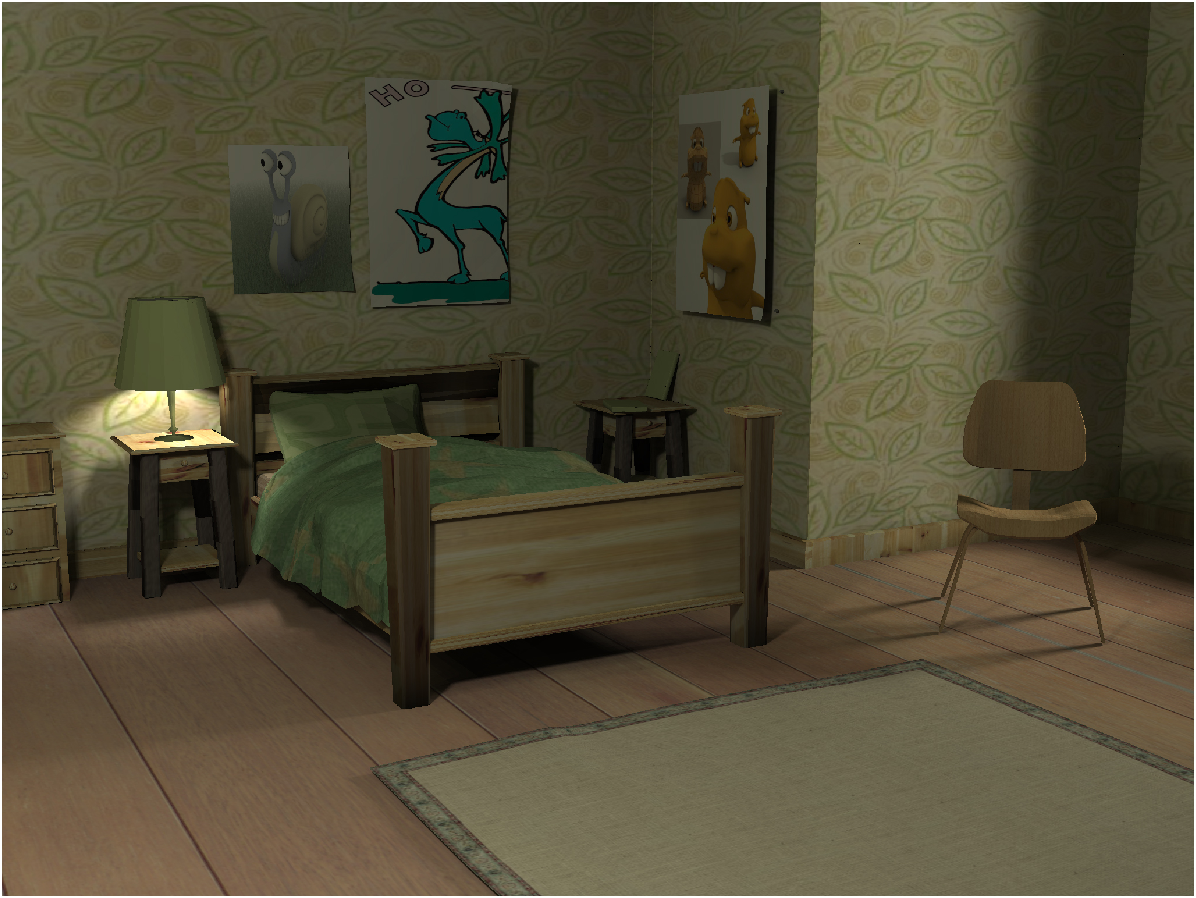} &
     \includegraphics[width=2.2in]{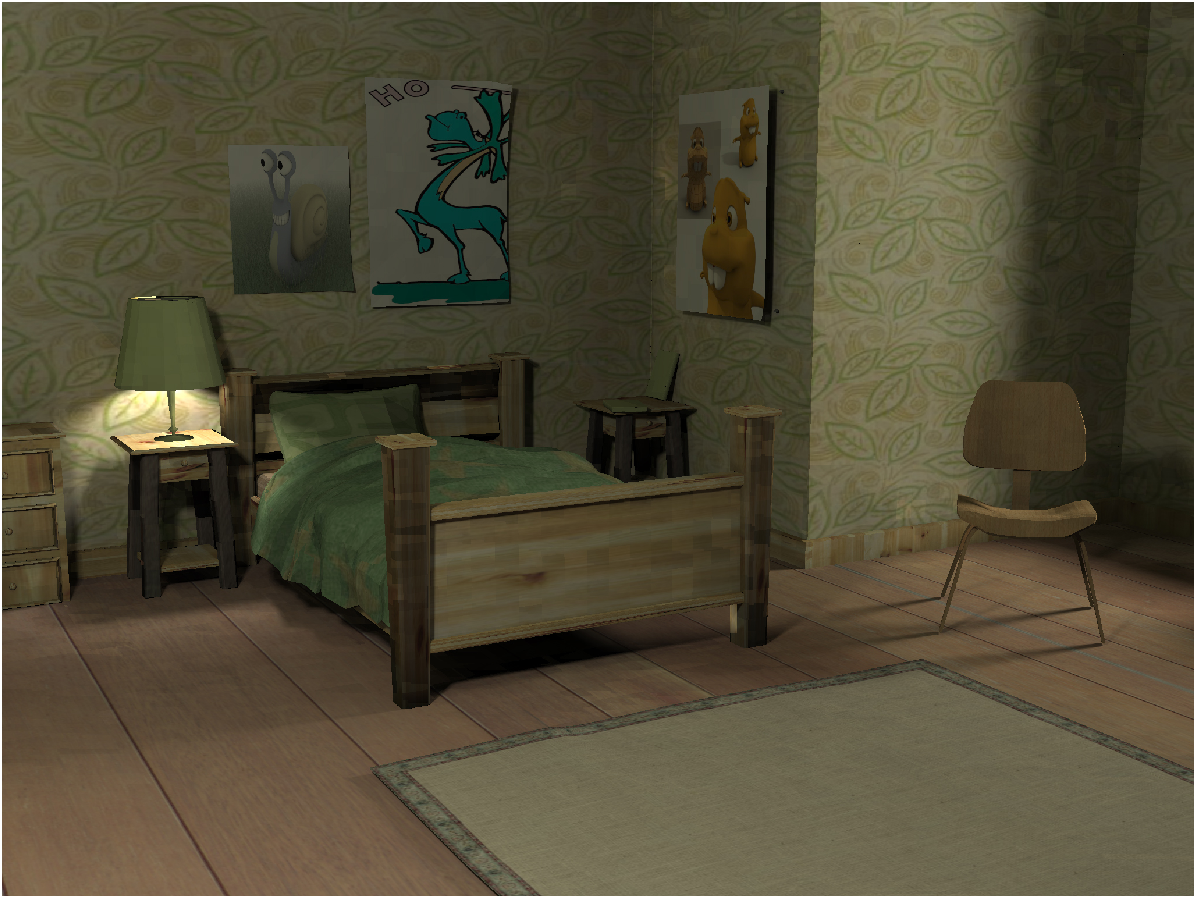} &
     \includegraphics[width=2.2in]{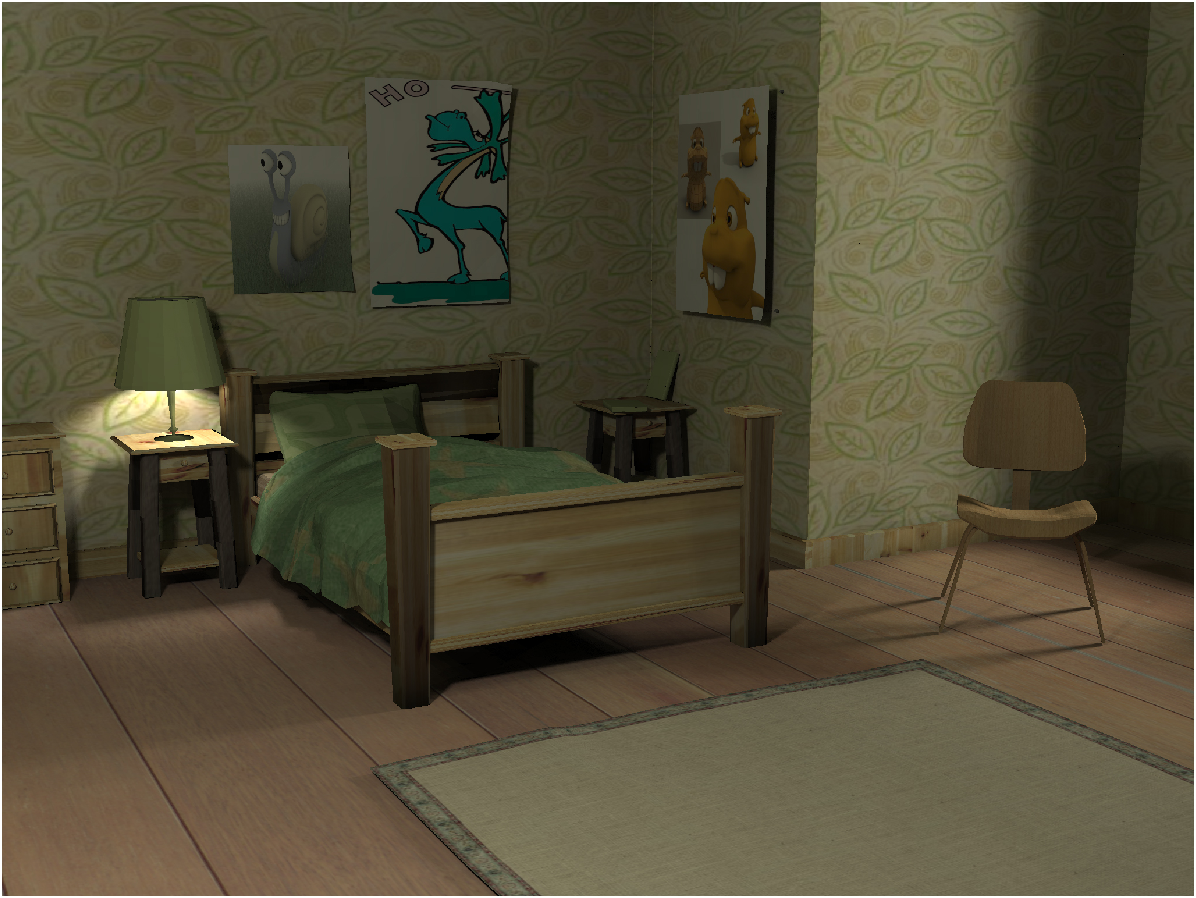} \\
     673 rays/pixel & 425 rays/pixel & 127 rays/pixel \\
     \includegraphics[width=2.2in]{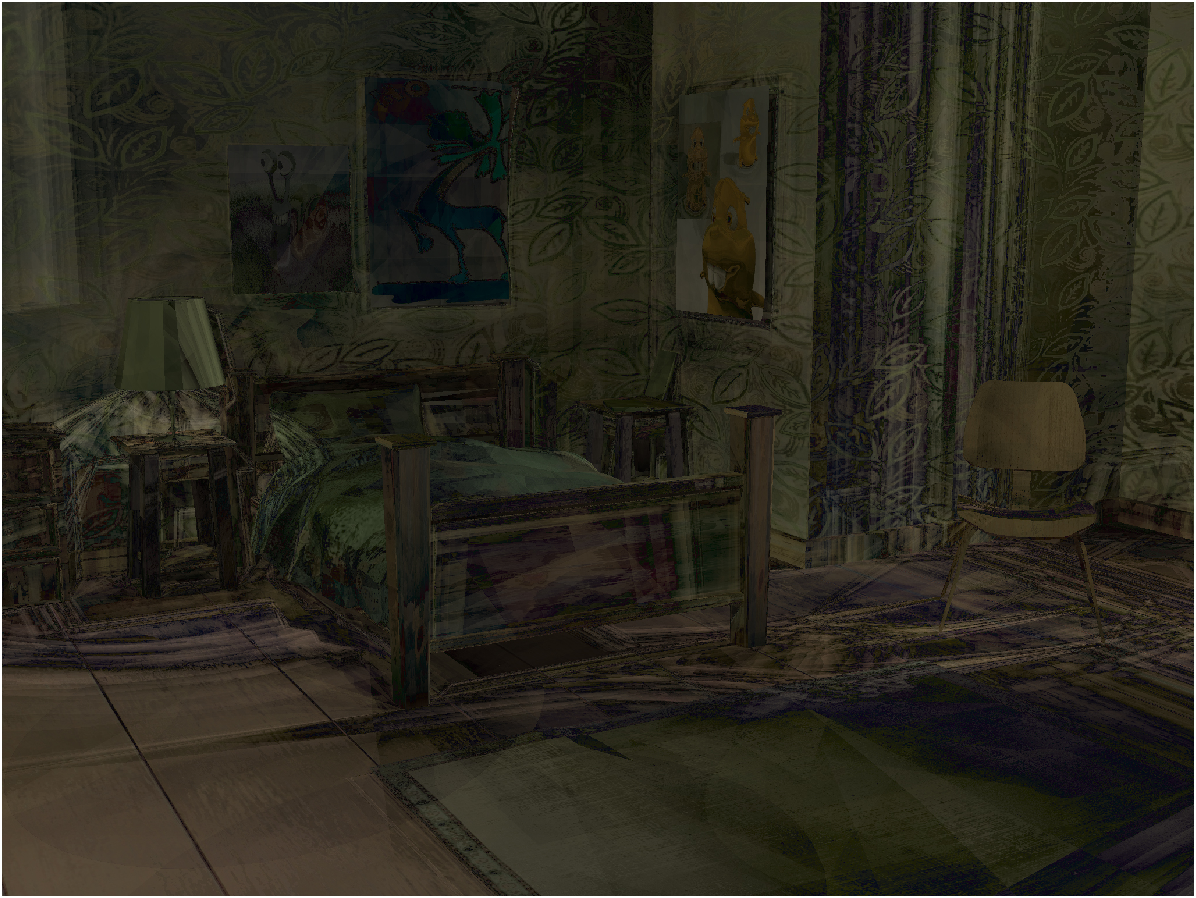} &
     \includegraphics[width=2.2in]{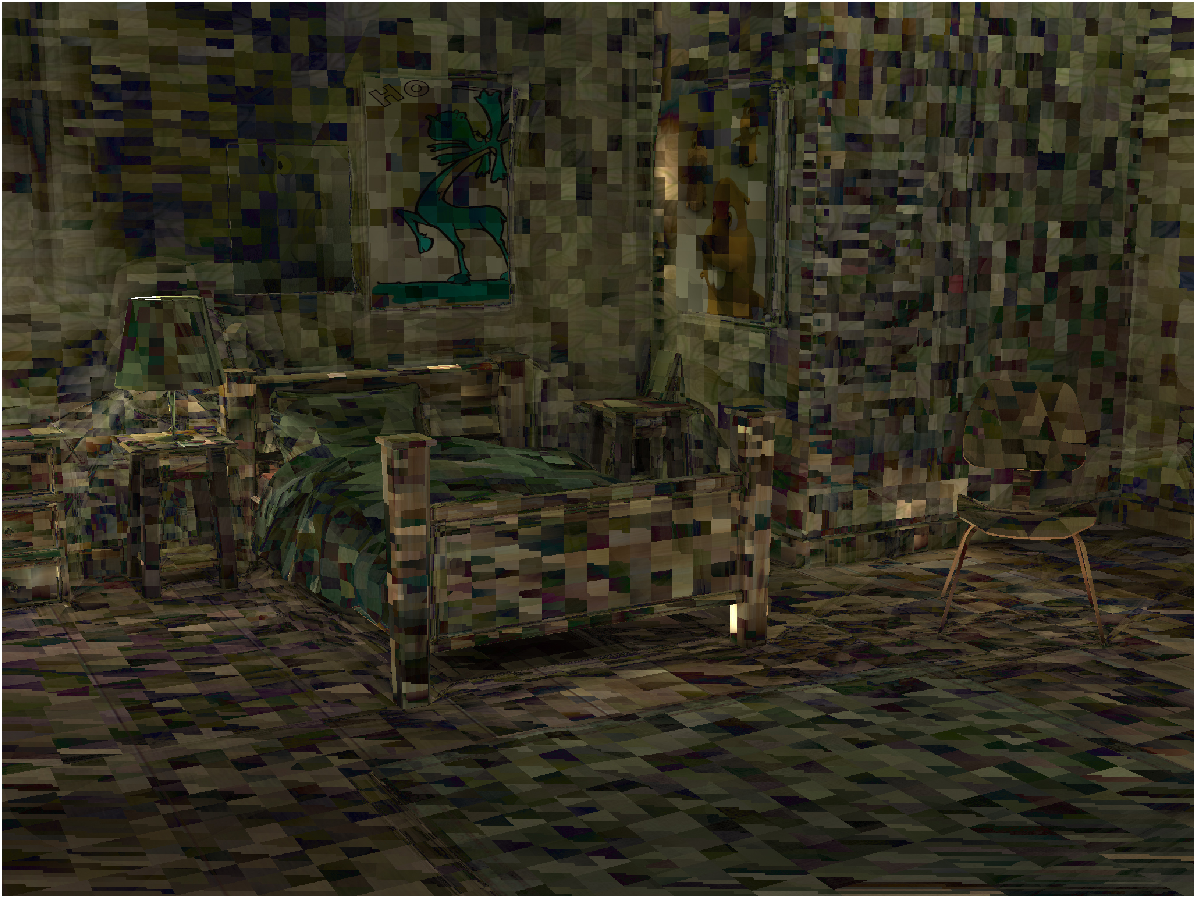} &
     \includegraphics[width=2.2in]{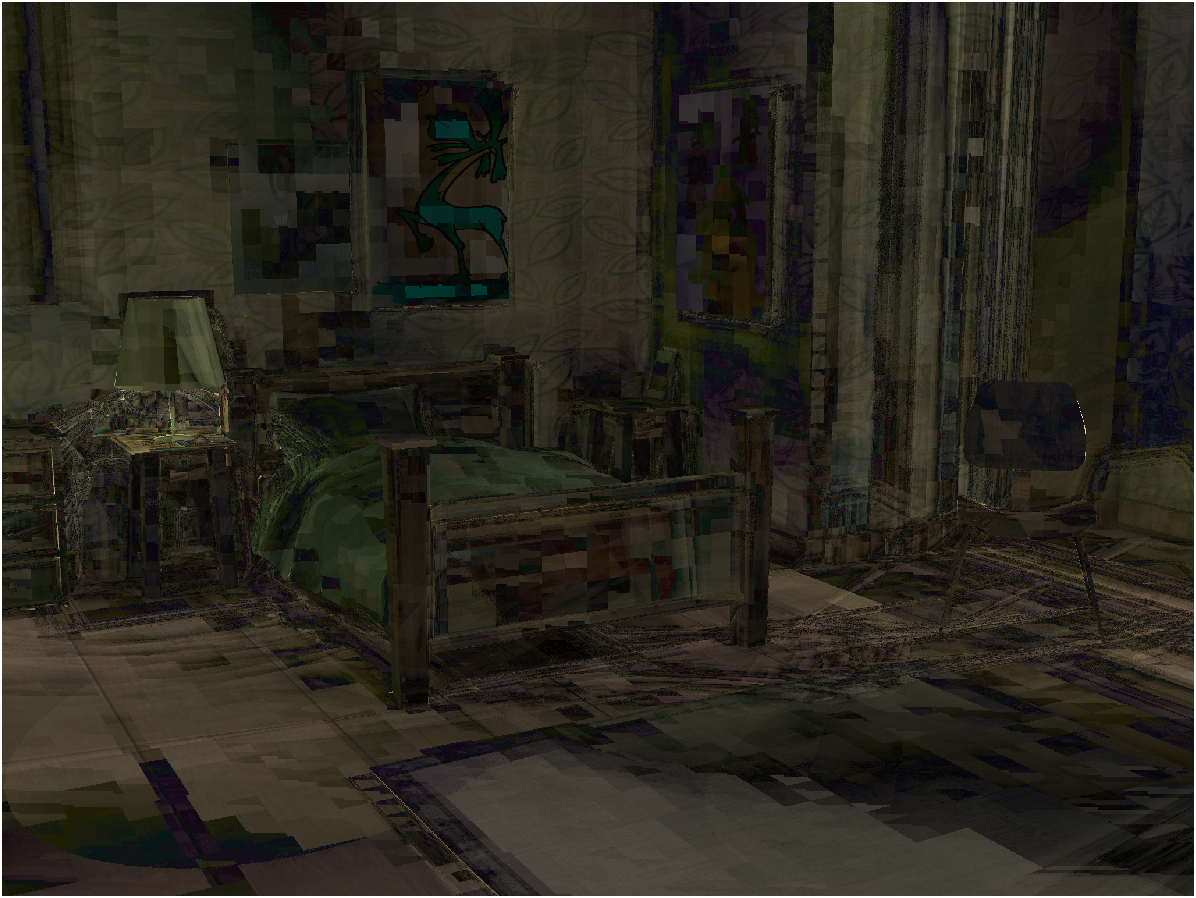} \\
     error=1.8\%  & error=2.8\%  & error=1.8\%  \\
     (a) lightcut & (b) lightslice & (c) ours \\
  \end{tabular}
} {The Bedroom scene rendered by various method. The bottom row shows the difference images multiplied by 8. }

\NewWideFigure{fig:f2} {
  \begin{tabular} {ccc}
     \includegraphics[width=2.2in]{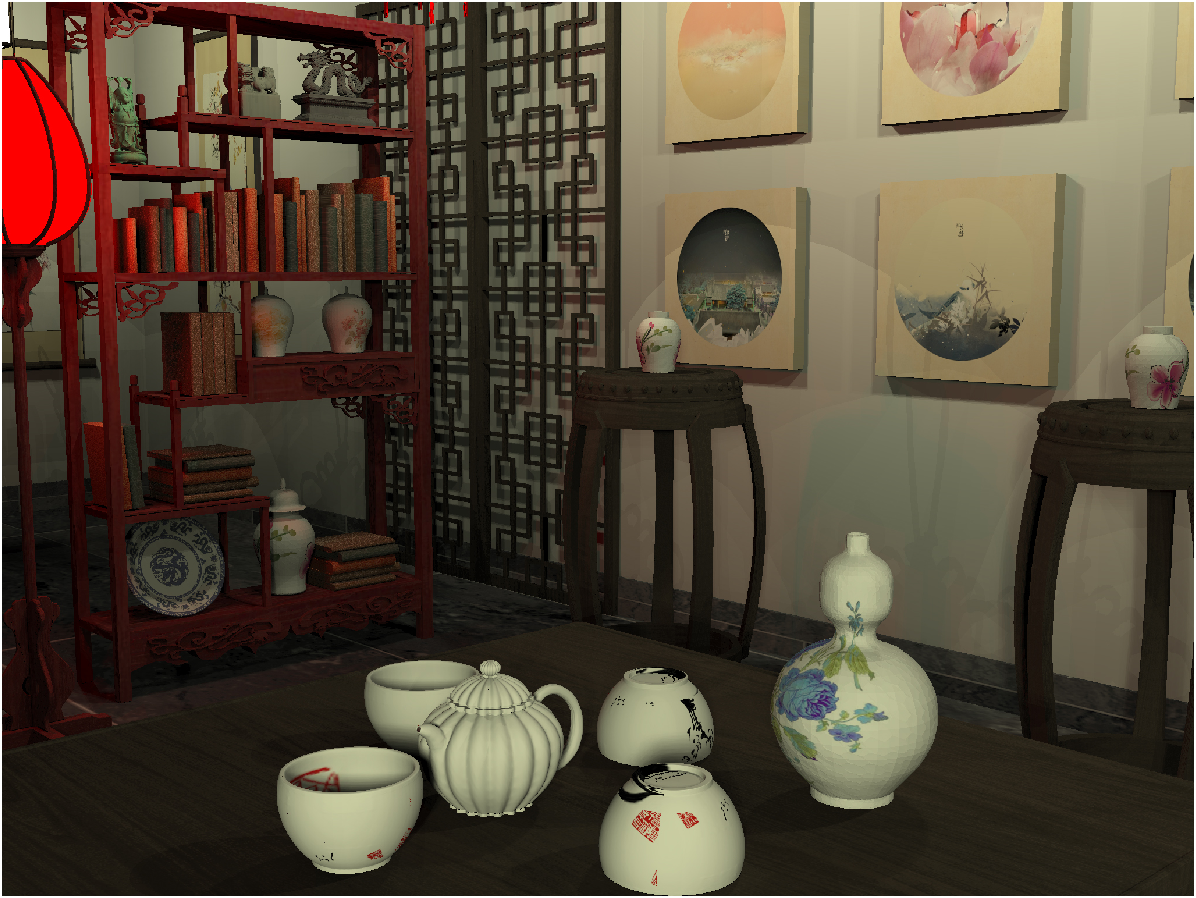} &
     \includegraphics[width=2.2in]{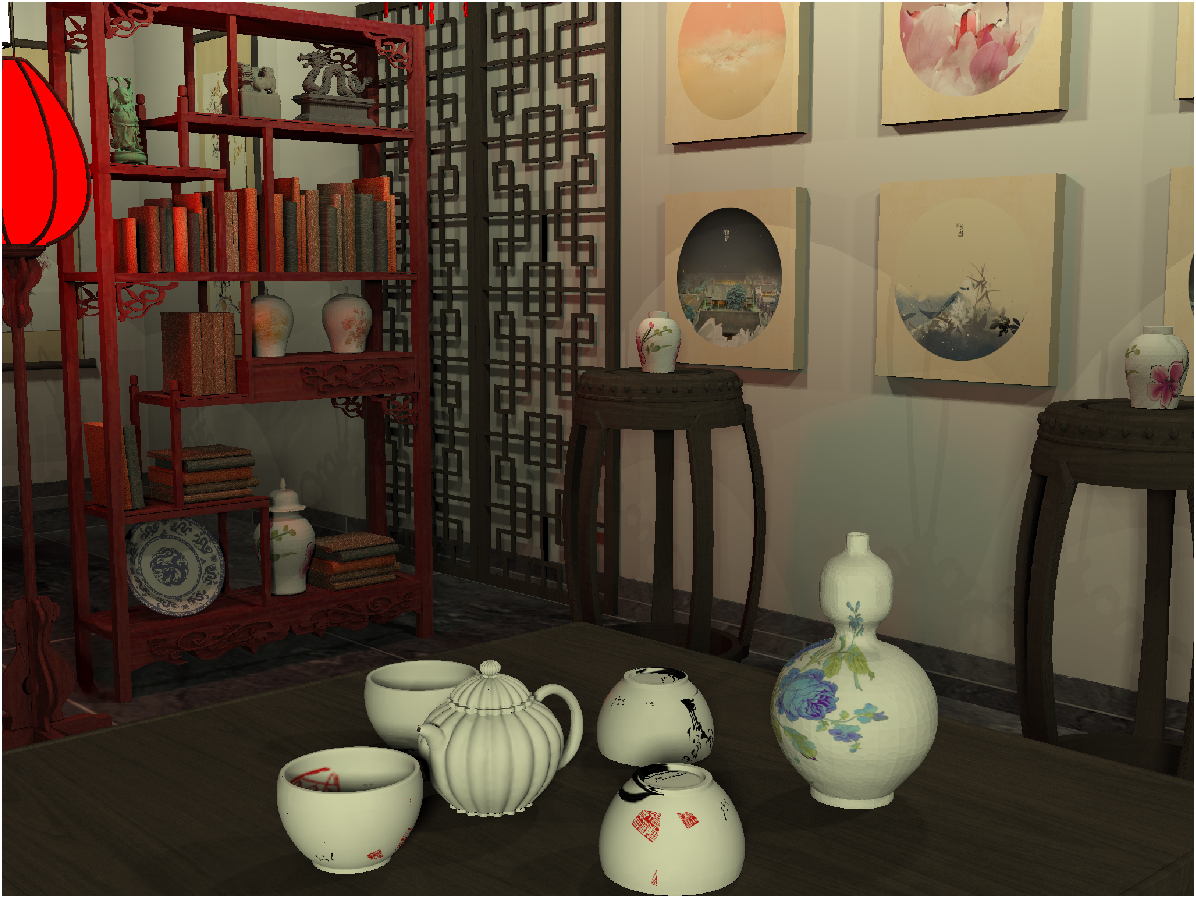} &
     \includegraphics[width=2.2in]{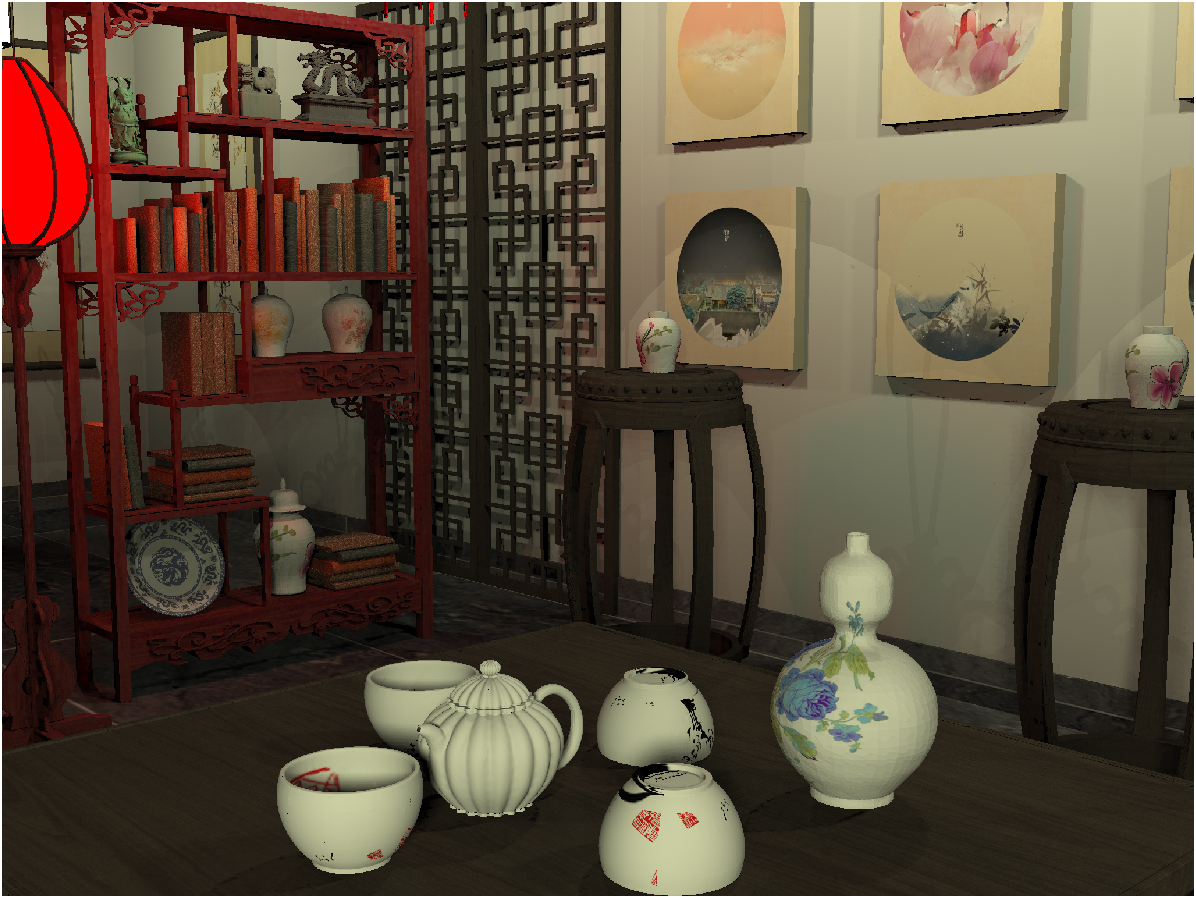} \\
      559 rays/pixel & 256 rays/pixel &  84 rays per pixel \\
     \includegraphics[width=2.2in]{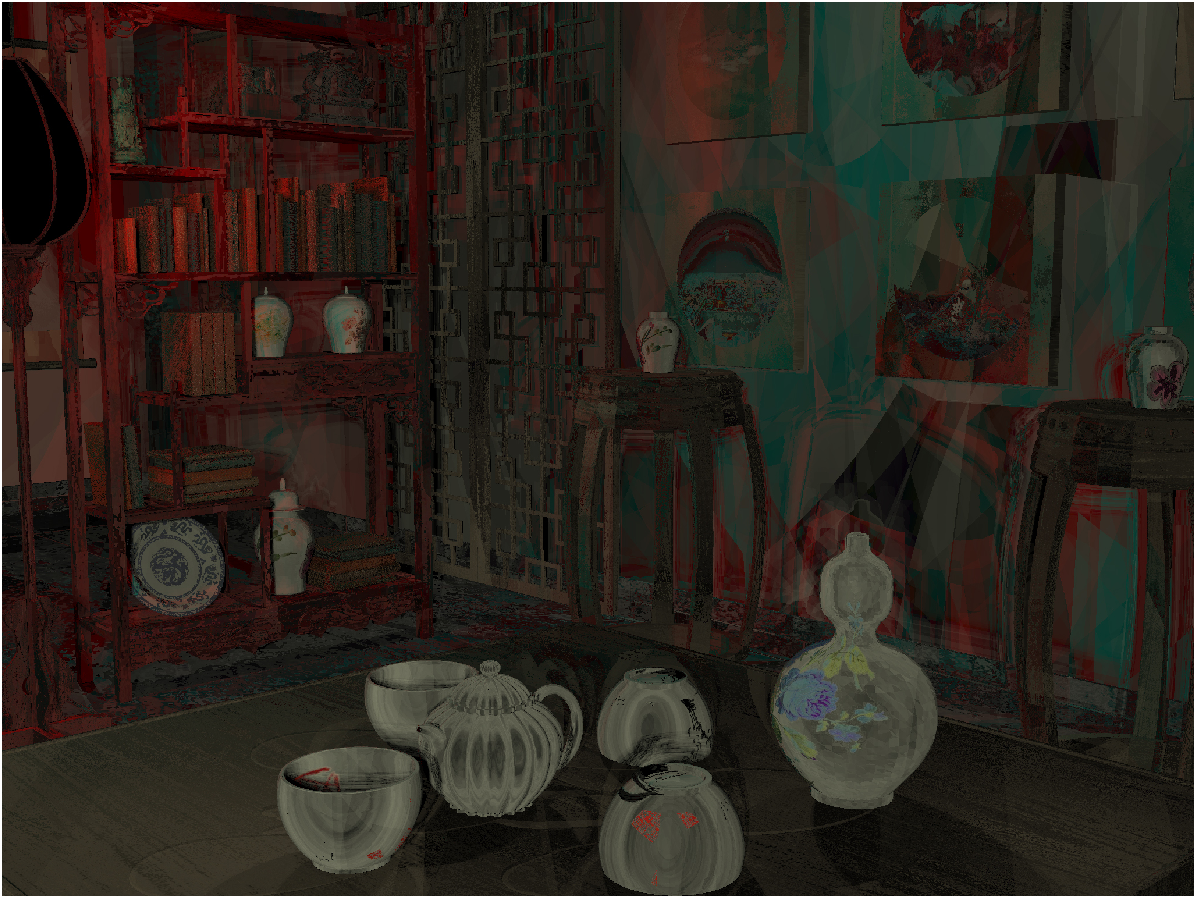} &
     \includegraphics[width=2.2in]{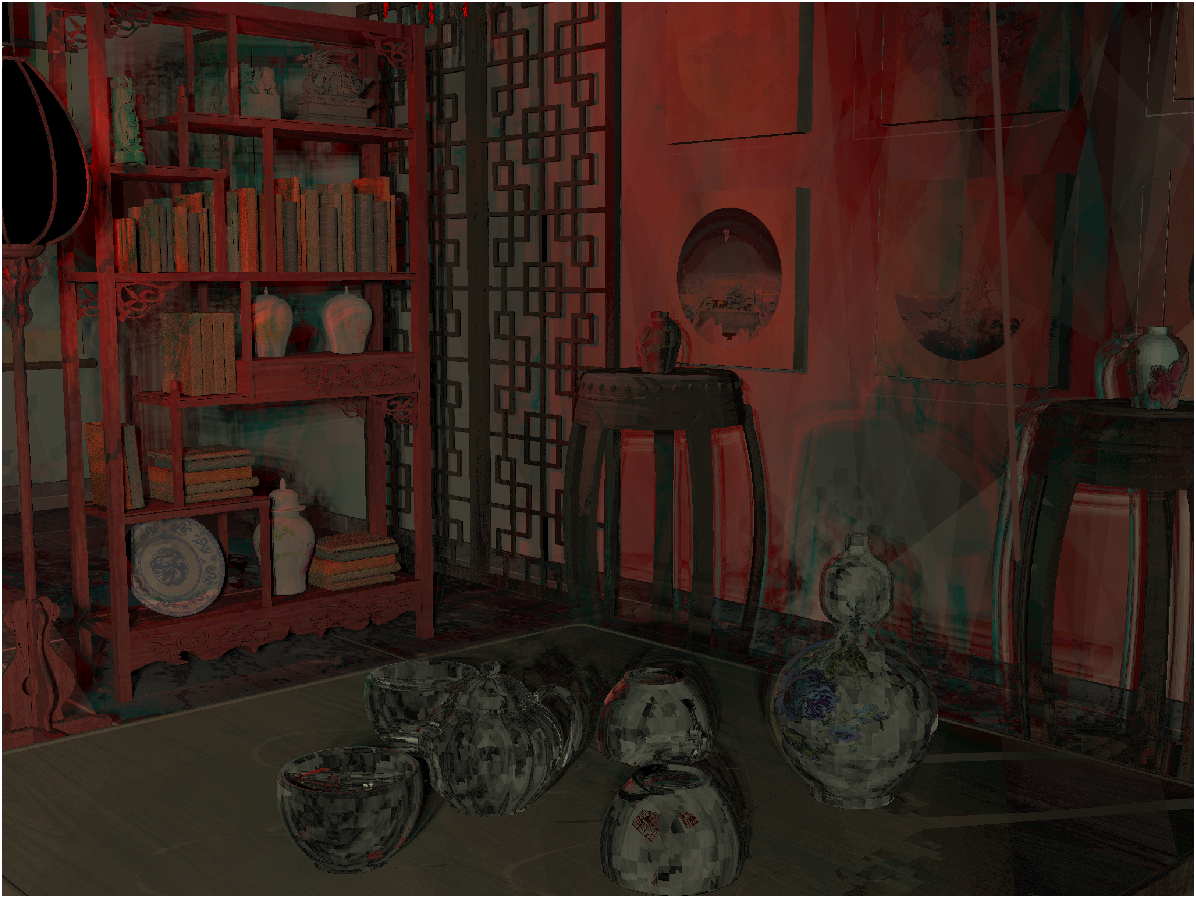} &
     \includegraphics[width=2.2in]{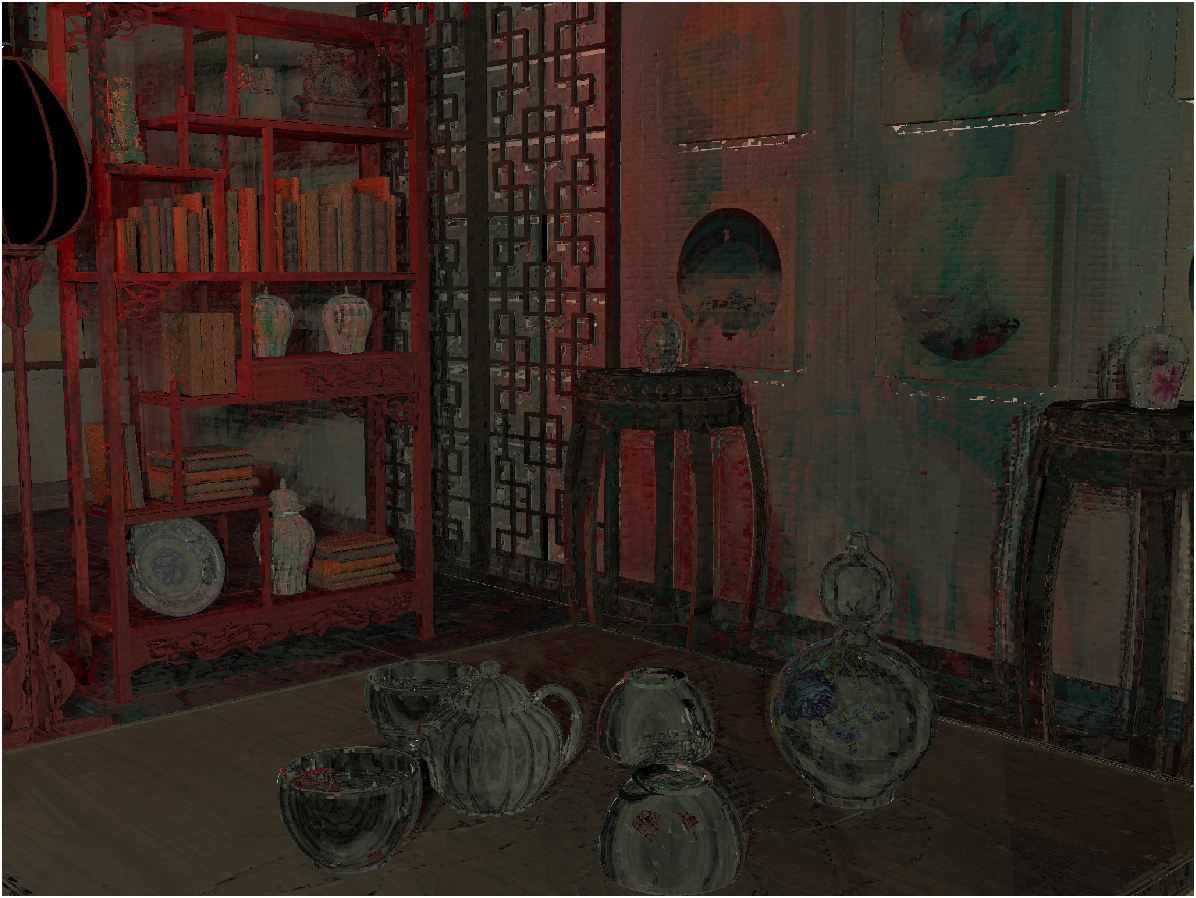} \\
     error=2.9\%  & error=3.3\%  & error=3.0\%  \\
     (a) lightcut & (b) lightslice & (c) ours \\
  \end{tabular}
} {The Tearoom scene rendered by various methods. The bottom row shows the difference images multiplied by 8. }

\refFig{fig:teaser} shows some results by changing the sampling rate in the step of matrix sampling and completion. The results shows that $10\%$ sampling rate is sufficient to deliver faithfully rendering results. And extremely low sampling rate, $1\%$, can deliver a reasonable good result for preview. These experiments verify that the lighting matrix is low rank, and can be sparse sampled and completed.

\refFig{fig:museum} shows a large complex scene with heavy occlusions. For this scene, our rendering takes about 3GB GPU memory. \refFig{fig:f1} shows some rendering results of a bedroom under a smooth lighting environment. It shows the our method can produce the best result using the smallest number of sampling rays per pixel. In this example, our method takes $10\%$ sparsely sampled entries for matrix completion. The smooth transition of the soft shadows and shadings are well reconstructed. One the contrary, the side wood of the bed exhibits perceptible band artifacts in the images rendered by the lightslice method. The reference image of this scene is shown \refFig{fig:ref}.

\refFig{fig:f2} compares some rendering results of complex indoor scene illuminated by 24 area light sources. In this scene, there are many casted shadows on the wall, and there are many regions receiving very strong local lightings. With bounded error control, the lightcut method can well preserve edge of the hard shadow (shown in the middle \refFig{fig:f2}(a)). Our method can reconstruct correct light effects and complex shadows using only $10\%$ sampling rate. The soft shadow boundaries shown in the middle \refFig{fig:f2}(c) is due to the low-rank approximation. \refFig{fig:f2}(b) shows that the lightslice method may make mistakes on the shows because insufficient number of rows are sampled when refine the lightcut for the slices with complex strong local lighting.

\section{Conclusion and Discussion}
We have presented an efficient rendering algorithm for the many-light problem based upon the idea of lightcut. The presented algorithm is featured by sparse sampling and two novel techniques based on spare sampling. One technique is the coarsening method for generating lightcuts by sparsely sampling the lighting matrix, and the other is the completion algorithm for the lighting matrix with a few sparsely sampled entries. Thanks to the sparse sampling strategy, the presented algorithm achieves significant acceleration compared with the state of the art many-light rendering algorithms.

Compared with the original lightcut method \cite{Walter:lightcuts05}, the success of our coarsening method owns to using the sparsely sampled visibility information. Since the original lightcut overestimated the visibility, it wastes many node splitting budget on the occluded nodes, leading to insufficient splitting budget on the visible nodes and more rendering errors. Our methods avoid visibility overestimation by computing sparse visibility.

The row-column sampling method \cite{Hasan:MatrixRowColumn} also generates a global light cut using the visibility information. It randomly selects a set of pixels and computes the lighting result from each lights. Since it uses the full visibility information regarding all lights, the number of selected pixels is limited to several hundreds for speedup. The lightslice method \cite{Ou:2011:LightSlice} partitions the image space into slices and refines the global light cut for each slice. However, only 1 pixel per slice is considered for the refinement. Our method is similar to the lightslice method in using slices, but ours computes the visibility information of a set of pixels in each slice, which can better capture the lighting characteristics.

The major difference between our method and the previous lightcut based methods lies in the matrix sampling and completion step. Instead of computing the full lighting matrix for rendering, ours sparsely samples it and reconstructs it by matrix factorization.

 One limitation of our sparse sampling and factorization is the lack of coherence, leading to flicking artifacts when moving the light and/or the camera. One possible direction to address this issue is to take advantage of the idea in multidimensional lightcut \cite{Walter:lightcuts06}. As another future work, it is interesting to exploit some image space features to guide the sampling for the lighting matrix factorization, while the pixels are uniformly random sampled in our current implementation,

\section*{Acknowledgements}

\section*{Appendix A -- Nonnegative Matrix Factorization}
Let $M\in R^{m\times n}$ be a nonnegative matrix, and $\Omega$ be a set of entry indices at which the value of matrix $M$ is known. The nonnegative factorization of $M$ is defined as minimizing the following Frobenius norm restricted on the known entries:
\begin{equation}
arg \min_{X,Y} \left\|\mathcal{P}_\Omega (XY-M) \right\|^2_F, \nonumber
\end{equation}
where $X \in R^{m\times q}$ and $Y \in R^{q\times n}$, and $X,Y>0$.
To use the alternating direction method (ADM) \cite{xu:11:alternating}, the above form is converted into the following equivalent form:
\begin{eqnarray}
arg \min_{U,V,X,Y,Z} \frac{1}{2}\left\|XY-Z \right\|^2_F \nonumber \\
s.t. \left\{\begin{array}{l}
      X=U, Y=V, \\
      U\geq0, V\geq0, \\
      \mathcal{P}_\Omega(Z-M)=0
     \end{array}\right. \nonumber
\end{eqnarray}
Then, applying ADM for the augmented Lagrangian in above, we have the following successive minimization steps:
\begin{equation}
\begin{array}{rcl}
  X_{k+1} & = &  (Z_k Y_k^T+\alpha U_k-\Lambda_k)(Y_k Y_k^T+\alpha I)^{-1},\\
  Y_{k+1} & = &  (X_{k+1}^T X_{k+1}+\beta I)^{-1}(X_{k+1} Z_k^T+\beta V_k-\Pi_k),\\
  Z_{k+1} & = &  X_{k+1}Y_{k+1}+\mathcal{P}_\Omega(M-X_{k+1}Y_{k+1}),\\
  U_{k+1} & = &  \mathcal{P}_+(X_{k+1}+\Lambda_k/\alpha),\\
  V_{k+1} & = &  \mathcal{P}_+(Y_{k+1}+\Pi_k/\beta),\\
  \Lambda_{k+1} & = & \Lambda_{k+1} + \gamma\alpha(X_{k+1}-U_{k+1}),\\
  \Pi_{k+1} & = &  \Pi_{k} + \gamma\beta(Y_{k+1}-V_{k+1}),\\
\end{array} \nonumber
\end{equation}
where $\Lambda\in R^{m\times q}$ and $\Pi \in R^{q\times n}$ are Lagrangian multipliers, $\alpha>0$ and $\beta>0$ are penalty parameters, $\gamma\in(0,1.618)$ is another parameter, operator $\mathcal{P}_\Omega$ set the entries at $\Omega$ to be 0s, and operator $\mathcal{P}_+$ clamps the negative entries by 0.

\bibliographystyle{ieee}
\bibliography{srbib}

\end{document}